\documentclass[useAMS, usenatbib]{mn2e}
\usepackage{color}
\usepackage{times}
\usepackage{natbib}
\usepackage{graphicx}
\usepackage{amsmath}
\newcommand{\logl}{{\rm log}L}

\newcommand{\oii}{\rm [O~{\textsc {ii}}]}
\newcommand{\oiii}{\rm [O~{\textsc {iii}}]}
\newcommand{\mbii}{MBII}
\newcommand{\hetdex}{HETDEX}
\newcommand{\gama}{GAMA}
\newcommand{\desi}{DESI}
\newcommand{\pgadget}{\textsc {p-gadget}}

\newcommand{\msun}{\rm M_\odot}
\newcommand{\halpha}{\rm H\alpha}
\newcommand{\pegasetwo}{{\sc pegase}.2}
\newcommand{\wfirst}{WFIRST-AFTA}

\title[LF of {[OII]} ELGs in the MBII simulation]
{Luminosity function of [OII] emission-line galaxies in the MassiveBlack-II
simulation}
\author[K. Park, T. Di Matteo, S. Ho, R. Croft, S. Wilkins, Y. Feng and N. Khandai]{KwangHo
Park$^{1,2}$\thanks{E-mail: kwangho.park@physics.gatech.edu (KP); tiziana@phys.cmu.edu (TDM);
shirleyh@andrew.cmu.edu (SH) }, 
Tiziana Di Matteo$^{1\star}$, Shirley Ho$^{1\star}$, Rupert Croft$^{1}$, 
\newauthor Stephen M. Wilkins${^3}$, Yu Feng$^{1,4}$ and Nishikanta Khandai$^{5,6}$ \\ 
$^{1}$McWilliams Center, Carnegie Mellon University, Pittsburgh,
PA 15213, USA \\ 
$^{2}$Center for Relativistic Astrophysics, Georgia Institute of
Technology, Atlanta, GA 30332, USA  \\
$^{3}$ Astronomy Centre, Department of Physics and Astronomy, University of Sussex, Brighton BN1 9QH, UK \\
$^{4}$ Berkeley Center for Cosmological Science, CA 94720, USA \\
$^{5}$ Brookhaven National Laboratory, Department of Physics,
Bldg 510, Upton, NY 11973, USA \\
$^{6}$ National Institute of Science Education and Research, Sachivalay Marg, PO Sainik School, Bhubaneswar-751005, Odisha, India}

\begin{document}
\date{Accepted. Received 2015}

\pagerange{\pageref{firstpage}--\pageref{lastpage}} \pubyear{2015}

\maketitle

\label{firstpage}

\begin{abstract}
We examine the luminosity function (LF) of [OII]~emission-line
galaxies in the high-resolution cosmological simulation MassiveBlack-II
(\mbii). 
From the
spectral energy distribution of each galaxy, we select a sub-sample
of star-forming galaxies at $ 0.06 \le z \le 3.0$ using the
[OII]~emission line luminosity L([OII]). 
We confirm that 
the specific star formation rate
matches that in the \gama~survey. We show that the [OII]~LF
at $z = 1.0$ from the \mbii~shows a good agreement with the LFs
from several surveys below L([OII])=$10^{43.0}$ erg\,s$^{-1}$ while
the low redshifts ($z \le 0.3$) show an
excess in the prediction of bright [OII] galaxies, but still
displaying a good match with observations below L([OII])=$10^{41.6}$
erg\,s$^{-1}$. Based on the validity in reproducing the properties
of [OII] galaxies at low redshift ($z \le 1$), we forecast the evolution of the
[OII]~LF at high redshift ($z \le 3$), which can be tested by
upcoming surveys such as the \hetdex~and \desi. The slopes of the
LFs at bright and faint ends range from -3 to -2 showing minima at
$z=2$. The slope of the bright end evolves approximately as $(z+1)^{-1}$
at $z \le 2$ while the faint end evolves as $\sim 3(z+1)^{-1}$ at $0.6
\le z \le 2$. In addition, a similar analysis is applied for the
evolution of [OIII]~LFs, which is to be explored in the forthcoming
survey \wfirst. Finally, we show that the auto-correlation function
of [OII] and [OIII] emitting galaxies shows a rapid evolution from
$z=2$ to $1$.
\end{abstract}


\section{Introduction}
\label{sec:intro}
Understanding the star formation history of the Universe is essential
in decoding the evolution and mass assembly of galaxies
\citep{LillyLHC:96,Madau:1996,Hopkins:2004,Hopkins:2006}. Observations
of star-forming galaxies at high redshift reveal that the cosmic
star formation rate (SFR) density has been decreasing since reaching
the peak at $z \sim 2$ \citep{Madau:1996, Hopkins:2006, KennicuttE:12}.
Thus, interpreting the evolution of the global SFR around this
redshift ($z \sim 2$) has been one of the main interests in the
astrophysics community. In particular, the role of self-regulated
processes of star formation and the effect of feedback by accretion
on to supermassive black holes (SMBH) at the centres of galaxies
have been the main clues in deciphering the observed star formation
history.

The SFR of an individual galaxy is estimated using several indicators
such as nebular emission lines or the UV continuum. Both indicators are
directly linked to the ionizing flux from young massive stars, whereas
the flux in the mid-infrared (MIR) or far-infrared (FIR) are used
to trace the re-radiated emission by dust.
Among the star formation indicators, $\halpha$ is regarded as one
of the best direct indicators of the current SFR since the line
strength of $\halpha$ is a good tracer of the photo-ionized gas
around massive young stars ($ < 20$~Myr)
\citep{Kennicutt:1998}. However, due to the rest-frame wavelength
($\lambda_{\halpha} = 6563$~\AA), it is challenging to use
$\halpha$ for galaxies at high redshift ($z > 1$) since the
observed wavelength is redshifted to a wavelength longer than NIR
($\lambda_{\rm obs} > 1~\micron$), which is difficult to
observe from the ground.
In contrast, $\oii~\lambda \lambda 3726, 3729$ emission line doublet
can still be observed from the ground in the wavelength $\lambda_{\rm
obs} < 1~\micron$ for galaxies at high redshift ($z < 1.6$) 
\citep{Gallego:02, Ly:07, Gilbank:2010, Zhu:2009, Ciardullo:2013,
Comparat:2015}. Since the $\oii$ line luminosity L([OII]) is an indirect
indicator (unlike $\halpha$), it requires an extensive calibration
based on the properties of the galaxies such as the metallicity and
stellar mass ($M_*$) \citep{KewleyGJ:2004, Moustakas:2006,
Gilbank:2010}. At even higher redshift ($z > 2$), one can use 
the UV continuum, which also directly traces the ionizing photons,
but it measures the SFR over longer timescales ($\sim 100$~Myr).

Over the past few decades, systematic searches for emission-line
galaxies (ELGs) have been made. \citet{Gilbank:2010} explored the
SFR at $z \sim 0.1$ using \oii, $\halpha$~and u-band luminosities
from the deep 275 deg$^2$ Stripe 82 field in the Sloan
Digital Sky Survey (SDSS) coupled with UV data from the Galaxy
Evolution EXplorer (GALEX) satellite. Also, the pilot survey of the
Hobby-Eberly Telescope Dark Energy Experiment
\citep[\hetdex;][]{Adams:2011,Ciardullo:2013} observed 284 \oii~emitting
galaxies at $z < 0.56$ in 169 arcmin$^2$. The \hetdex~pilot survey
is a blind integral-field spectroscopic study of four data-rich 
areas of sky: COSMOS \citep{Scoville:2007}, GOODS-N
\citep{Giavalisco:2004}, MUNICS-S2 \citep{Drory:2001}, and XMM-LSS
\citep{Pierre:2004}. The Galaxy And Mass Assembly
survey~\citep[\gama;][]{Bauer:2013} also targets $\sim 73,000$
galaxies at $ 0.05<z<0.32$ with $M_* < 10^{10}~\msun$. The \gama~survey
calculates the SFR using the $\halpha$ emission line and explores how
the specific SFR (sSFR) depends on the stellar mass of the galaxies.
Recently, \citet{Comparat:2015} explored the evolution of the
luminosity function (LF) of the \oii~emitters in the redshift range
$0.1 < z < 1.65$ based on the medium-resolution flux-calibrated
spectra of ELGs with the VLT/FORS2 instrument and the SDSS-III/BOSS
spectrograph. In the forthcoming years, there are ambitious surveys
targeting ELGs at high redshift. The Dark Energy Spectroscopic
Instrument (\desi) is a multi-fiber spectrograph that will target
18 million ELGs \citep{Schlegel:09,Levi:2013}. \desi~aims to probe
the effect of the dark energy on the expansion history of the
Universe using baryon acoustic oscillations (BAO) with a sky coverage
of 14,000--18,000\,deg$^2$. Another survey is the space-based
Wide-Field InfraRed Survey Telescope-Astrophysics Focused Telescope
Assets (\wfirst)~which will perform a deep infrared survey of
$2400$\,deg$^2$ and obtain spectroscopic redshifts of galaxies
targeting 20 million $\halpha$~emitters at $1<z<2$ and 2 million
\oiii~ELGs at $2<z<3$ \citep{Spergel:13}.

In cosmological simulations, it is challenging to study stellar
mass assembly of individual galaxies as it requires a high
resolution and a large volume at the same time.  
A sufficiently large volume is needed for statistically meaningful results
whereas high resolution properly manifests baryonic physics
related to star formation and black hole (BH) feedback. It is also difficult to
evolve a simulated universe up to the present ($z \sim
0$). Considering all these numerical challenges, the {\it
MassiveBlack-II} simulation \citep[\mbii;][]{Khandai:2014},
which is a successor to the {\it MassiveBlack} simulation
\citep{DiMatteo:2012}, is one of the unique simulations in that 
it is a state-of-the-art high resolution cosmological
hydrodynamical simulation with a large comoving volume
\citep[c.f.,][]{Vogelsberger:2014, Schaye:2015}. 

In this paper, we investigate the properties of the star-forming
galaxies in the \mbii~simulation, which are selected by the \oii~and
\oiii~emission-line luminosity in the generated spectral energy 
distributions (SEDs). The main
goal of this study is to validate whether the star-forming galaxies
in the high-resolution cosmological simulation can reproduce results
from recent surveys for the \oii~emitting galaxies and make a prediction 
for the upcoming surveys that target the high redshift \oii~emitters.
In Section~\ref{sec:method} we explain how we select
the \oii~emitting galaxies from the synthesised SEDs and compare
the properties of the selected samples with observations. In
Section~\ref{sec:results}, we compare the LFs from the simulation
and observation at $0.1 \le z \le 1.0$, and make a prediction for
the LF of the \oii~emitting (and \oiii) galaxies at high redshift.
We also present the evolution of the auto-correlation function of
the \oii~ELGs at $z \le 4 $. Finally, we summarise and discuss the
results in Section~\ref{sec:summary}.

\section{Methods}
\label{sec:method}
\subsection{The MassiveBlack-II~simulation}
We use the state-of-the-art high resolution hydrodynamical simulation
\mbii~\citep{Khandai:2014} performed with the hybrid version
TreePM-SPH code \pgadget~designed for running on \textsc{petaflop}-
scale supercomputer facilities. 
The \mbii~contains $N_{\rm part}=2
\times 1792^3$ dark matter and gas particles in a comoving volume
of $(100~h^{-1}{\rm Mpc})^3$ to satisfy the needs of large volume
and high resolution at the same time. In the \mbii~simulation, the
$\Lambda$CDM universe is evolved from $z = 159$ to $z =0.06$ with
a high mass resolution of $1.1\times10^7~h^{-1}\msun$ for dark
matter and $2.2\times10^6~h^{-1}\msun$ for gas particles with a
smoothing length of $1.85~h^{-1}$kpc. The cosmological
parameters used in the \mbii~are: $\sigma_8=0.816$, spectral index
$n_s=0.968$, matter
density fraction of the critical density $\Omega_{\rm m}=0.275$,
vacuum energy density fraction $\Omega_\Lambda=0.725$, baryon density
fraction $\Omega_{\rm b}=0.046$, and Hubble constant $h=0.702$ from
the 7-year Wilkinson Microwave Anisotropy Probe \citep{Komatsu:2011}.


The \mbii~includes a sub-grid model for star-forming multi-phase
gas \citep{Springel:2003}. In this model, a thermal instability
takes place at a critical density threshold creating a multi-phase
medium consisting of cold clouds in pressure equilibrium with 
surrounding hot gas. A star formation prescription is given by the
Kennicutt-Schmidt law \citep{Kennicutt:1989}, where the SFR is
proportional to the density of cold clouds (i.e., $\rho_{\rm SFR} \propto
\rho_{\rm gas}^{N}$ where $N=1.5$ is adopted). Gas particles are
converted to star particles according to the star formation
prescription. Star formation is then regulated by the supernovae
feedback, which heats the surrounding gas and creates a self-regulated
cycle.

In the \mbii, a BH is introduced as a collisionless sink particle
in a newly collapsing halo identified by the friends-of-friends
halo-finder on-the-fly at a regular time interval. A seed BH
with mass $M_{\rm seed}= 5\times 10^5~h^{-1}\msun$ is inserted into
a halo with mass $M_{\rm halo} \ge 5 \times 10^{10}~h^{-1}\msun$.
After seeded, BHs are assumed to grow at the Bondi-Hoyle rate
\citep{BondiH:44,Bondi:52} $\dot{M}_{\rm BH} =  4 \pi G^2
\rho_\infty M_{\rm BH}^2 (c_{s,\infty}^2+v_{\rm BH}^2)^{-3/2}$, where
$v_{\rm BH}$ is the velocity of the BH relative to the surrounding
gas, and $\rho_\infty$ and $c_{s, \infty}$ are the density and sound
speed of the gas in the multi-phase state \citep{Pelupessy:07}.
The accretion rate is limited by $2\times \dot{M}_{\rm Edd}$
\citep{BegelmanVR:06, Volonteri:2006} where $\dot{M}_{\rm Edd}$ is
the Eddington accretion rate. The accretion rate is
converted to a bolometric luminosity $L_{\rm bol} =\eta \dot{M}_{\rm
BH} c^2$, where $\eta$ is the radiative efficiency and we adopt the
standard value of $0.1$ for a thin disk model \citep{ShakuraS:73}.
It is also assumed that 5 per cent of the BH luminosity thermally
couples with the surrounding gas, isotropically depositing the
radiation energy to the gas particles within the BH kernel (64
nearest neighbours) to match the observed $M_{\rm BH}$--$\sigma$
relation \citep{DiMatteo:05,Springel:05}. The current BH growth
model has been adopted in extensive studies
\citep{Li:2007,Sijacki:2007,DiMatteo:2008,Colberg:2008,
Croft:2009,Degraf:2010,Degraf:2011a,Degraf:2011b,
DiMatteo:2012,DeGraf:2012,Chatterjee:2012,Feng:2014}. BHs are assumed
to merge when one BH enters within the kernel of another BH with a
relative velocity below the local gas sound speed.

\subsection{[OII]~emission-line galaxies in the \mbii}
The high resolution of the \mbii~enables us to generate a SED of an 
individual galaxy. We adopt the previous work by \citet{Wilkins:2013b} 
to generate the SEDs and we briefly explain the process in this section.

\begin{figure}
\includegraphics[width=85mm]{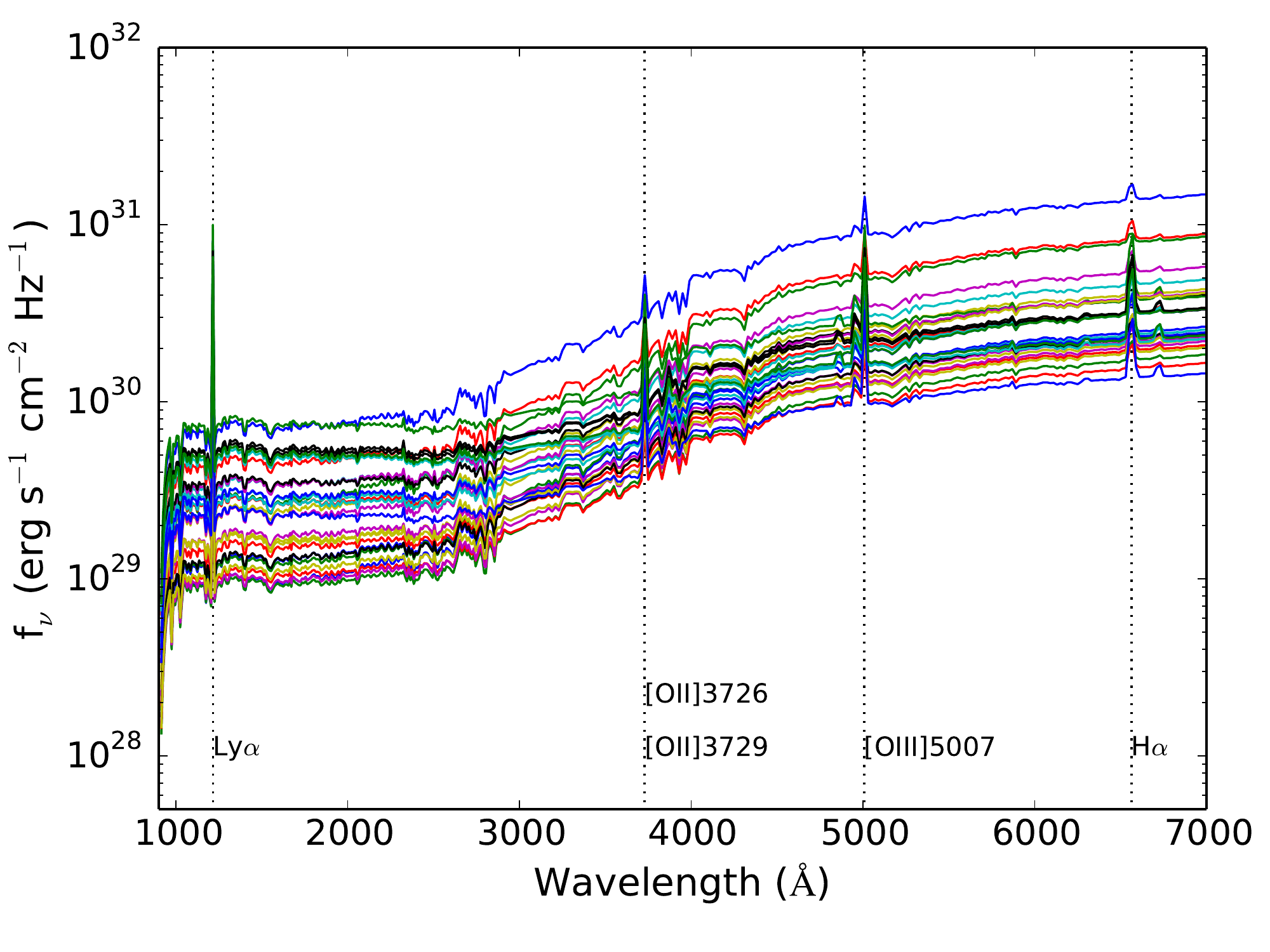}
\caption{Examples of the spectral energy distribution of 20 star-forming
galaxies in the range $10 < $ SFR ($\msun$ yr$^{-1}$) $<100$ at $z=0.1$ in
the \mbii~simulation. The stellar continuum and nebular emission
lines are generated from the star formation history of each galaxy
using the stellar population synthesis code \pegasetwo. Note that
the nebular emission lines such as Ly$\alpha$,
\oii$~\lambda \lambda $3726, 3729, \oiii~$\lambda$5007, and H$\alpha$
are distinct.} 
\label{fig:high_sf_seds}
\end{figure}

\subsubsection{Generating SEDs with emission lines}
The stellar population synthesis code \pegasetwo~\citep{Fioc:1997,
Fioc:1999} is used to generate SEDs of individual star
particles as a function of the stellar mass, age, and metallicity
and assuming the Salpeter initial mass function.
We then calculate hydrogen line fluxes first using the
\pegasetwo~while non-hydrogen line fluxes are estimated from hydrogen
lines using the metallicity-dependent conversion by \citet{Anders:2003}.
These nebular emission lines, reprocessed ionizing radiation by
interstellar medium, include Ly$\alpha$, \oii$~\lambda \lambda$3726,
3729, \oiii$~\lambda$5007, and H$\alpha$.
The rest-frame SEDs of galaxies are generated by integrating all
the SEDs of star particles and emission lines
with a wavelength resolution of
$20$~\AA~\citep{Wilkins:2013b, Wilkins:2013a}. For a galaxy with a
stellar mass $M_* = 10^9~h^{-1} \msun$, for example, $\sim$\,450 SEDs from
individual star particles are integrated to produce the 
SED of the galaxy. Fig.~\ref{fig:high_sf_seds} shows 20 randomly selected
examples of synthesised SEDs of star-forming galaxies with $10 < $
SFR ($\msun$ yr$^{-1}$) $<100$ at $z=0.1$. Note that the nebular emission
lines, such as Ly$\alpha$, \oii, \oiii, and H$\alpha$, are clearly
visible for these star-forming galaxies.

\subsubsection{Selecting [OII]~emission-line galaxies from SEDs}
We select a sample of \oii~ELGs using the generated
SED for each galaxy. The flux of \oii~emission line for an individual galaxy
is calculated from the SED by mimicking observation. We apply a
simple criterion L$_{\rm \lambda 3730}$/L$_{\rm cont} > 1$, where
L$_{\rm cont}$ is the estimated flux of the continuum at $\lambda
3730$ by averaging the fluxes at the neighbouring wavelengths. This
method selects galaxies with positive flux at the \oii~emission line
wavelength. Then the luminosity of the \oii~emission line component
is obtained by subtracting the estimated continuum as L(\oii) =
L$_{\rm \lambda 3730}-$ L$_{\rm cont}$. We check that this simple
method recovers the original line \oii~luminosity well (see
Appendix~\ref{appendix:lf_raw}).

\begin{figure}
\includegraphics[width=85mm]{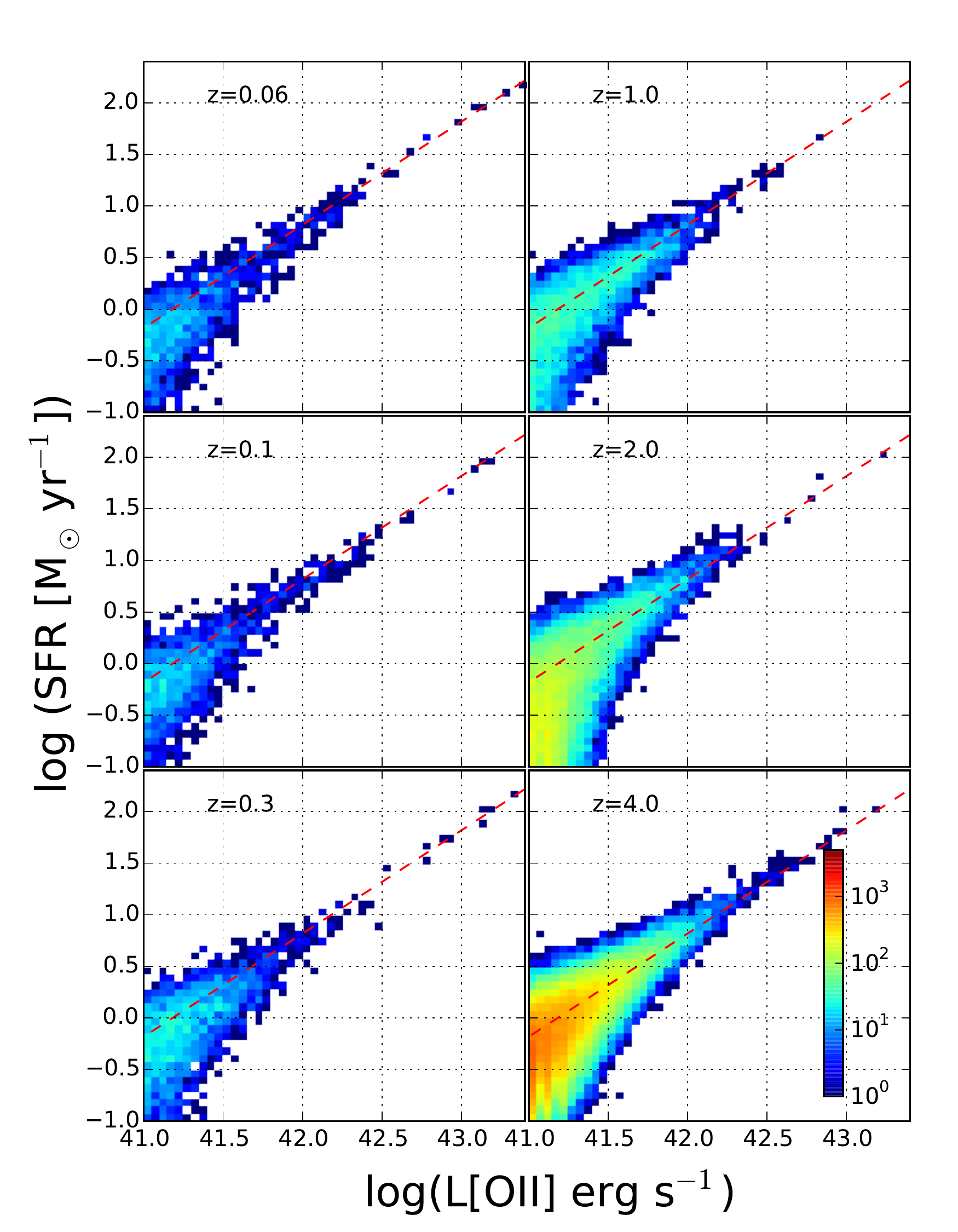}
\caption{2D histogram of the \oii~luminosity and the SFR in the
\mbii~for different redshifts ($z=$0.06, 0.1, 0.3, 1.0, 2.0 and
4.0). The dashed line in each panel shows the empirical relationship
from \citet{KewleyGJ:2004}, which shows a good agreement with the
\mbii~simulation. The galaxies with higher SFR show a tight match
while the scatter increases as the L(\oii)~decreases.} 
\label{fig:oii_sfr} 
\end{figure}

\subsubsection{Comparison with L([OII])--SFR empirical relation}
Fig.~\ref{fig:oii_sfr} shows the relation between L(\oii) and SFR
of the \oii~ELGs at redshifts $z =$ 0.06, 0.1,
0.3, 1.0, 2.0, and 4.0. The \mbii~simulation displays a good agreement
with the empirical relationship by \citet{KewleyGJ:2004},  

\begin{equation} 
{\rm SFR}~(\msun~{\rm yr}^{-1}) = \frac{L(\oii)}{1.52\times10^{41}}  
({\rm erg}~s^{-1}), 
\label{eq:sfr_oii}
\end{equation}
which is shown as a dashed line in each panel. In general, the
galaxies with higher SFRs show a better agreement with the empirical
relation while the scatter increases as L(\oii) decreases. Due to
the mass resolution of star particles in the \mbii~(i.e., $m_* =
2.2 \times 10^6~h^{-1}\msun$), shot noise starts to dominate for
the sample of galaxies with L(\oii) $\le 10^{40.6}$~erg\,s$^{-1}$
(see Appendix~\ref{appendix:lf_raw} for details).
The current SFR is not well
represented in L(\oii) below this luminosity cut, which sets the
lower limit of L(\oii) for our current study. This luminosity cut
can be translated to a SFR of $\sim$\,0.3\,$\msun$\,yr$^{-1}$ using
equation~(\ref{eq:sfr_oii}).


\begin{figure}
\includegraphics[width=85mm]{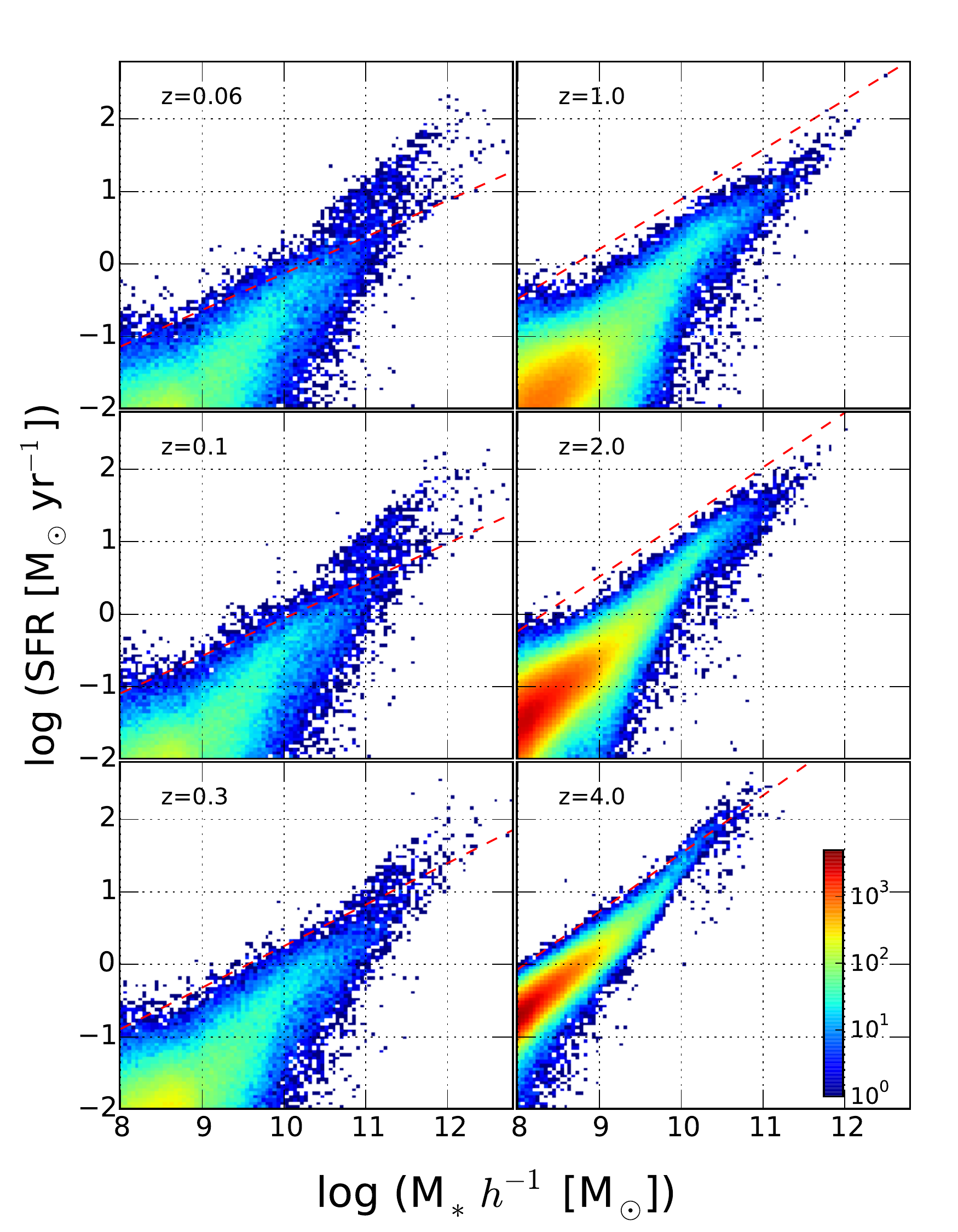}
\caption{2D histogram for the distribution of SFR and stellar mass
($M_*$) of galaxies in the \mbii~at the redshift $z=$0.06, 0.1,
0.3, 1.0, 2.0 and 4.0, respectively. Most of the star-forming
galaxies with $M_* \ge 10^{9}~h^{-1} \msun$ at $z=0.06$ are not
selected by the L(\oii) while most of the star-forming galaxies
with the same stellar mass at $z=4.0$ are selected. The
dashed line for each redshift is from \citet{Speagle:2014}.}
\label{fig:smass_sfr} 
\end{figure}

Fig.~\ref{fig:smass_sfr} shows the distribution of the stellar mass
($M_*$) and SFRs of all star-forming galaxies in the \mbii~for
redshifts $z=$0.06, 0.1, 0.3, 1.0, 2.0 and 4.0.  In general, the
SFR is proportional to the stellar mass $M_*$ for each redshift, and
the SFR for a given stellar mass increases as we go to high redshift.
Note that the number of stellar particles is $\sim$\,450 for a galaxy
with stellar mass $M_* \sim 10^{9}~h^{-1} \msun$. When the luminosity
cut mentioned above is applied (i.e., SFR of $\sim 0.3~\msun$~yr$^{-1}$),
it removes a large fraction of galaxies at low stellar mass while
the sample of star-forming galaxies is more complete for massive
galaxies. The average stellar mass of galaxies selected
this way is approximately $M_* \ge 10^{10}~h^{-1} \msun$ 
at $z=0.06$. The completeness at a given stellar mass also depends
on redshift. For example, most of the star-forming galaxies with
$M_* \sim 10^{9}~h^{-1} \msun$ at $z=0.06$ are not selected for
our luminosity cut while most of the star-forming galaxies with the
same stellar mass are selected at $z=4.0$. 

For comparison, the observed
relation between the SFR and the galaxy stellar mass from \citet{Speagle:2014}  

\begin{equation}\begin{split}
&{\rm log\,SFR}(M_*, t) = (0.84 \pm 0.02 - 0.026 \pm 0.003 \times t) {\rm log}\,M_* \\
&-(6.51 \pm 0.24 - 0.11 \pm 0.03 \times t)
\end{split}\end{equation}
where $t$ is the age of the universe in Gyr, is overplotted in
Fig.~\ref{fig:smass_sfr}. Since it is known that the
\mbii~produces too many galaxies at low and high stellar mass 
\citep{Khandai:2014}, it is expected that the overabundance in
stellar mass function inevitably propagates into the \oii~LF. 
The excess is obviously seen in Fig.~\ref{fig:smass_sfr}
at low redshifts. 

\subsubsection{Specific star formation rate}
\begin{figure}
\includegraphics[width=85mm]{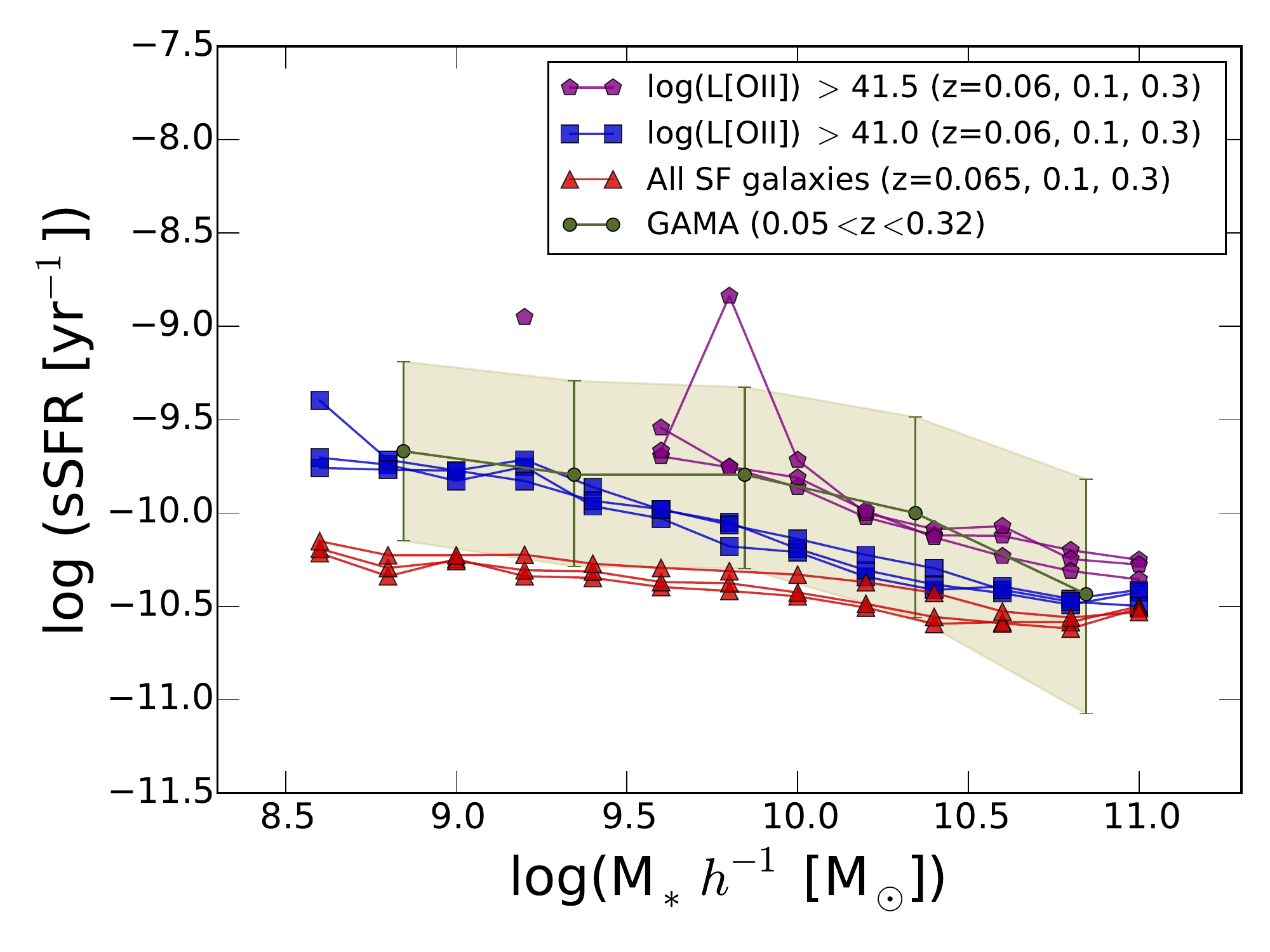}
\caption{Specific star formation rate (sSFR) as a function of the
stellar mass ($M_*/\msun$) at $z=0.06, 0.1$, and $0.3$. Different
symbols are used for different luminosity cuts: triangles are all
star-forming galaxies, squares are galaxies with L(\oii) $\ge
10^{41.0}$~erg\,s$^{-1}$, and pentagons are galaxies with
L(\oii) $\ge 10^{41.5}$~erg\,s$^{-1}$. The shaded region shows the
mean sSFR with scatter from the \gama~survey \citep{Bauer:2013} at
$0.05 < z < 0.32$ as a function of the stellar mass in the range
$9.0 <$ log~$(M_*~[\msun]) < 11.0 $.}
\label{fig:ssfr_gama}
\end{figure}

In this section, we compare the SFR per stellar mass (i.e., sSFR=SFR/$M_*$)
of the \oii~emitters with the \gama~survey \citep{Bauer:2013}. Fig.~\ref{fig:ssfr_gama}
shows the average sSFR as a function of the stellar mass in the
range $9.0<$ log$(M_*~[\msun])<11.0$ for redshifts
$z=$~0.06, 0.1, and 0.3
with different L(\oii) cuts.
Triangles are all star-forming galaxies, squares are galaxies
with L(\oii) $\ge 10^{41.0}$~erg\,s$^{-1}$, and pentagons are galaxies
with L(\oii) $\ge 10^{41.5}$~erg\,s$^{-1}$.  The shaded region in
Fig.~\ref{fig:ssfr_gama} shows the observed average sSFR with scatter
from the \gama~survey. The mean sSFR increases with higher luminosity cuts and
the sSFR with the luminosity cut of $10^{41.5}$~erg\,s$^{-1}$
matches well with the \gama~survey. However, this luminosity cut
removes galaxies with $M_* \le 10^{10}\,\msun$. The simulation
converges to the result of \gama~survey with increasing luminosity
cut, but the halos with stellar mass $M_* < 10^{10}~ \msun$
are lost with this luminosity cut (L(\oii) $\ge 10^{41.5}$~erg\,s$^{-1}$).

\begin{figure*}
\includegraphics[width=120mm]{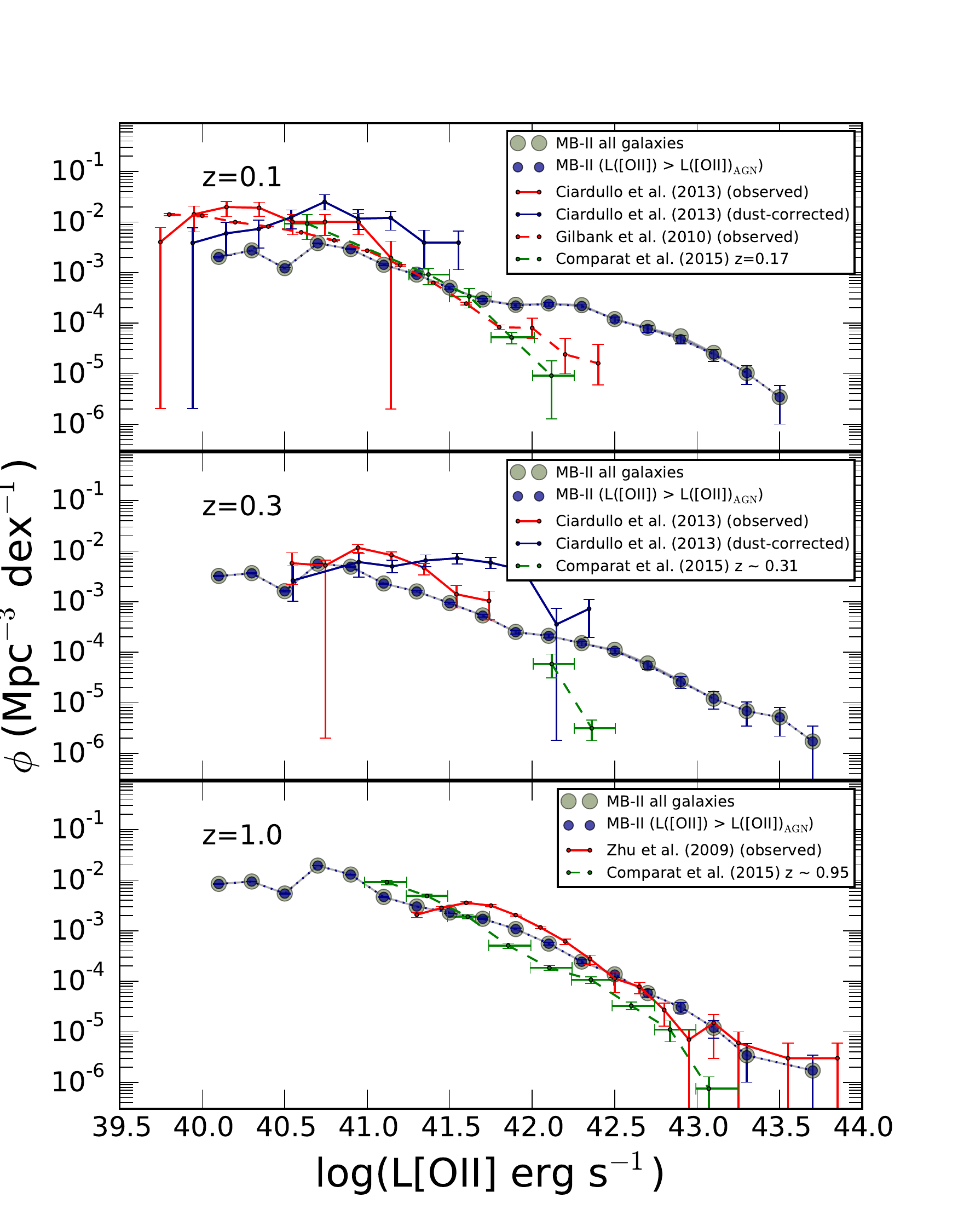}
\caption{Comparison of the LFs of the \oii~emitting galaxies in the
\mbii~simulation at $z=0.1$ (top), $0.3$ (middle), and $z=1.0$
(bottom) with the observations. Big circles show the \oii-selected
sample in the \mbii~simulation and the small circles along with the 
Poisson errors show the samples
with AGNs excluded using the method mentioned in Section~\ref{sec:agn}
to avoid the possible contamination of AGNs. There is no severe
contamination to the \oii~emission line. Red (observed) and blue
(dust-corrected) solid lines at $z=0.1$ and $0.3$ are from \hetdex~pilot
survey \citep{Ciardullo:2013} while the red dashed line (observed)
at $z=0.1$ shows the observation from SDSS \citep{Gilbank:2010}.
The LFs of the \mbii~show an excess in the prediction of bright
\oii~emitters at L(\oii) $>$ 10$^{41.6}$ erg\,s$^{-1}$ at
low redshift ($z \le 0.3$). At $z=1.0$, the LF matches well with
the DEEP2 observation \citep{Zhu:2009} and \citet{Comparat:2015}
below L(\oii)=$10^{43.0}$ erg\,s$^{-1}$.}
\label{fig:lumi_func} 
\end{figure*}

\subsubsection{Possible contamination from active galactic nuclei}
\label{sec:agn}
In our scheme, the SED of a galaxy does not include the emission
lines from the hot ionized gas produced by active galactic
nuclei (AGN). We assume that the whole \oii~emission line flux
originates from SFR. AGN contamination to \oii~line
certainly affects the estimation of the SFR
of a galaxy, but observations show that the level of actual
contamination is not high \citep[e.g.,][]{Bauer:2013, Zhu:2009}.

However, we exclude galaxies that might potentially contaminate our
samples when the AGN-associated \oii~emission line is added to the
current SEDs. We only select galaxies where the \oii~luminosity
from star formation is larger than that from the AGN as L$(\oii)_{\rm
AGN} \le$ L(\oii)$_{\rm SF}$. From the accretion rate of the SMBHs
for each galaxy, we calculate the bolometric luminosity of the AGN
as $L_{\rm AGN}=\eta \dot{M}_{\rm BH} c^2$ where $c$ is the speed
of light and we assume the radiative efficiency $\eta = 0.1$.
The direct relation between $L_{\rm AGN, bol}$ and L(\oii) is unclear,
so we use the results by \citet{Heckman:2004} and \citet{Netzer:2009}
where the \oiii $\lambda 5007$ is used as a proxy for the bolometric
luminosity of the AGN $L_{\rm AGN, bol}$ and the average relation is
shown as $L_{\rm AGN, bol}/L_{\rm AGN, \oiii} \sim 3500$. Since the
line ratio between \oii~and \oiii~in AGNs ranges from 0.1--5
depending on the types \citep{Blandford:1990}, we can roughly obtain
the fraction of the L(\oii) out of the bolometric luminosity as $3
\times 10^{-5} \le {\rm L}(\oii)_{\rm AGN}/L_{\rm AGN, bol} \le 0.001$.

\begin{figure*}
\includegraphics[width=180mm]{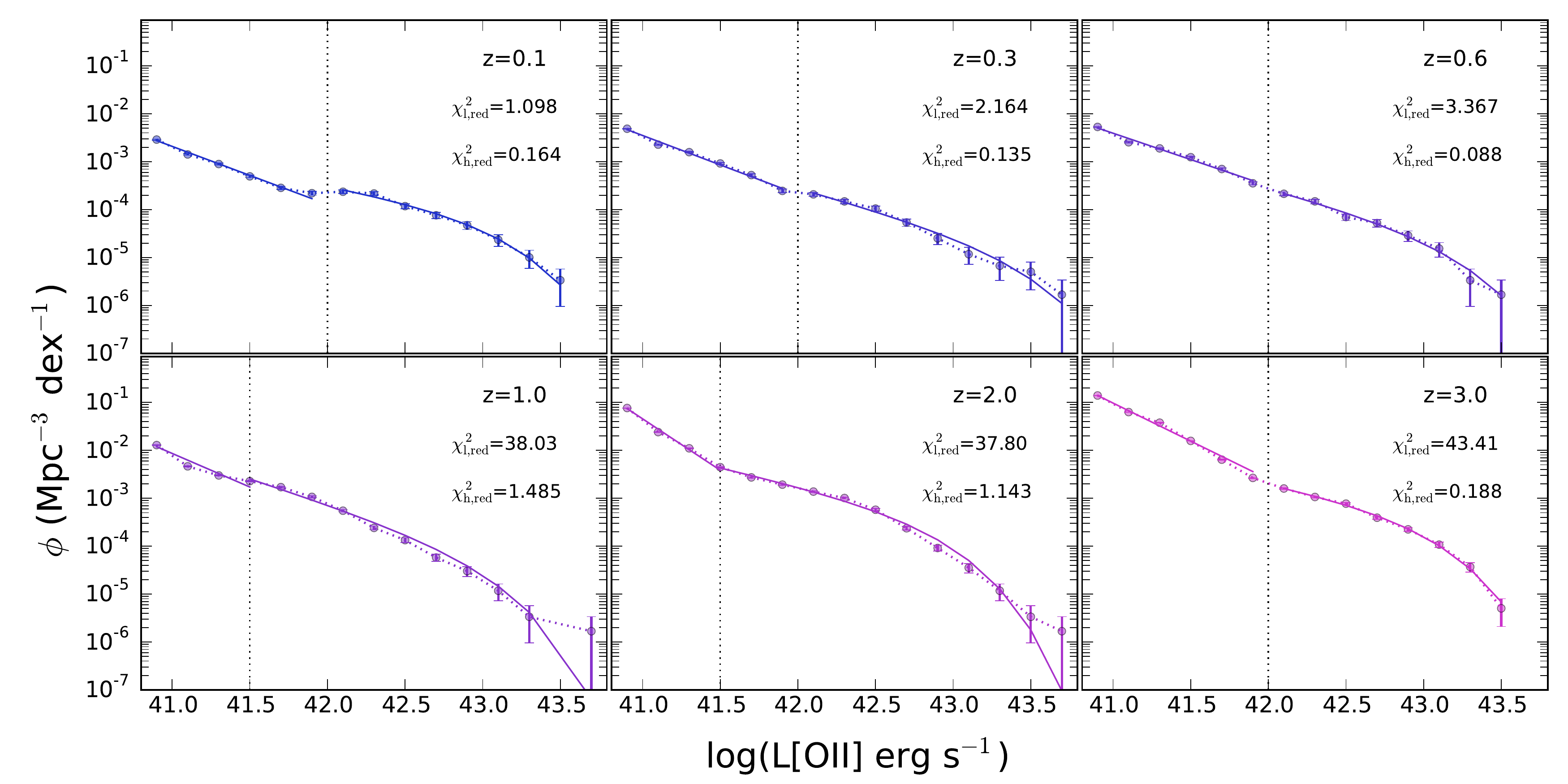}
\caption{Fitting of LFs of the \oii~emitting galaxies at $z=0.1$,
0.3, 0.6, 1.0, 2.0, and 3.0. For each redshift, we fit the LF with
a single power-law at low luminosity and Schechter function at high
luminosity. We use L(\oii) $= 10^{42.0}~$erg\,s$^{-1}$ as a dividing
luminosity of the fitting functions except for $z=1.0$ and $2.0$
where we apply L(\oii) $= 10^{41.5}~$erg\,s$^{-1}$, which are shown
as dotted vertical lines in each panel.
} 
\label{fig:fit_oii} \end{figure*}

\subsubsection{Dust extinction effect}
In this section, we briefly discuss the effect of the dust extinction
on the L(\oii) even though we rather use the intrinsic L(\oii)
without dust correction in this study. It is extremely
challenging to understand the effect of the dust extinction on
nebulae emission lines coupled with the star formation history,
metallicity, and the evolution as a function of redshift
\citep{Wilkins:2013b}.

Both the stellar continuum of the SEDs and emission lines are
affected by the dust extinction. In observation, the
internal dust extinction should be corrected to obtain the
intrinsic flux at the wavelength where the emission lines are located. 
On the simulation side, dust-extinction effect must be added to make
a direct comparison with observation. Since the wavelength
$\lambda=3730~{\rm \AA}$ is in $U$ band, we can apply 
$E(U-V)_{\rm star} = 1.64 E(B-V)_{\rm star}$ to the stellar continuum. 
In galaxies actively forming stars, the empirical
relationship of the internal dust extinction \citep{Calzetti:2001}
shows that the emission from the ionized gas suffers about twice
as much reddening as the stellar continuum $E(B-V)_{\rm star} =
0.44 E(B-V)_{\rm gas}$. Then the absorption in the $U$ band becomes
$A_{\rm U,gas} = A_{\rm V, gas} + (1.64 A_V)/(0.44 R_V)$. 
Applying the commonly accepted reddening coefficient $R_V \sim 3.1$
and the extinction value at $V$ band $A_{\rm V}=0.4~$mag from which
$A_{\rm V,gas}=0.91$ is obtained [e.g., $A_{\halpha} = 0.7^{+1.4}_{-0.7}$
for $z \sim 0.5$ galaxies \citep{Ly:2012b}], the extinction coefficient
of \oii~emission line is estimated as $A_{\oii} \sim 1.39~$mag.
Therefore, the dust absorption reduces the observed \oii~line
luminosity L(\oii)$_{\rm o}$ to $\sim 0.28$ times of the intrinsic
flux L(\oii)$_{\rm i}$.
Many studies find that the dust reddening $E(B-V)$ also depends on the
intrinsic \oii~luminosity L(\oii)$_{\rm i}$. For example,
\citet{KewleyGJ:2004} measure $E(B-V)=(0.174 \pm 0.035) {\rm
log~L}(\oii)_{\rm i} - 6.84$, so the intrinsic \oii~line luminosity
is expressed as L(\oii)$_{\rm i} = 3.11 \times 10^{-20}$L(\oii)$_{\rm
o}^{1.495}$ where the luminosities are given in unit of ergs per
second.

\section{Results}
\label{sec:results}
\subsection{[OII]~luminosity function at low redshift}
We compare the LFs of the \oii~emitting galaxies in the \mbii~simulation
with observations at redshift $z=$ 0.1, 0.3, and 1.0 from top to
bottom in Fig.~\ref{fig:lumi_func}. Circles show the LFs from the
\mbii~simulation while lines show the observational results
\citep{Zhu:2009,Gilbank:2010,Ciardullo:2013, Comparat:2015} at each
redshift. Big circles show all the \oii-selected galaxies while
small circles with the Poisson errors show the galaxies with AGNs
excluded to avoid a possible AGN contamination. We compare the
L(\oii) with the estimated L(\oii)$_{\rm AGN}$ due to AGN activity
for each galaxy using the method described in Section~\ref{sec:agn}.
We assume that L(\oii)$_{\rm AGN}$= 0.001$\times$L$_{\rm AGN, bol}$ to be
conservative. Note that the level of the AGN contamination in the
L(\oii) LF is negligible as discussed in Section~\ref{sec:agn}.

At $z=0.1$, red (observed) and blue (dust-corrected) solid lines
show \hetdex~pilot survey \citep{Ciardullo:2013} while the red
dashed line (observed) shows the result from SDSS \citep{Gilbank:2010}.
The \hetdex~pilot survey has a higher number density compared to SDSS
at $z=0.1$, but the excess can be explained by the cosmic
variance caused by the small sample size ($< 300$ galaxies at $0.1
< z < 0.56$) as discussed in \citet{Ciardullo:2013}. \citet{Comparat:2015}
matches well with other observations at low L(\oii) although it fits
better with a Schechter function when the high L(\oii) samples are
considered together. LFs from the \mbii~are a good match with the
observation \citep{Gilbank:2010,Comparat:2015} in the range of
\oii~luminosity 10$^{40.6}$ $<$ L(\oii)$ <$ 10$^{41.6}$~erg\,s$^{-1}$. 
The LFs of the \mbii~show an excess in the prediction of bright
\oii~emitters at L(\oii) $>$ 10$^{41.6}$ erg\,s$^{-1}$ at
low redshift ($z \le 0.3$). 
Note that the galaxy stellar mass function (GSMF) of
the \mbii~at low redshift is known to show an excess compared to
observations both in the low- and high-mass ends \citep{Khandai:2014}.
Since a high fraction of the star-forming galaxies in the high-mass
end is selected as \oii~emitting galaxies, the overabundance in
GSMF at the high-mass end is inevitably transferred to the \oii~LF.

At $z=1.0$, the LF of the \mbii~shows a good match with \citet{Zhu:2009}
throughout the luminosity that the DEEP2 survey covers while the LF 
matches with \citet{Comparat:2015} below L(\oii)=$10^{43.0}$
erg\,s$^{-1}$, as shown at the bottom panel of Fig.~\ref{fig:lumi_func}.

\begin{figure}
\includegraphics[width=85mm]{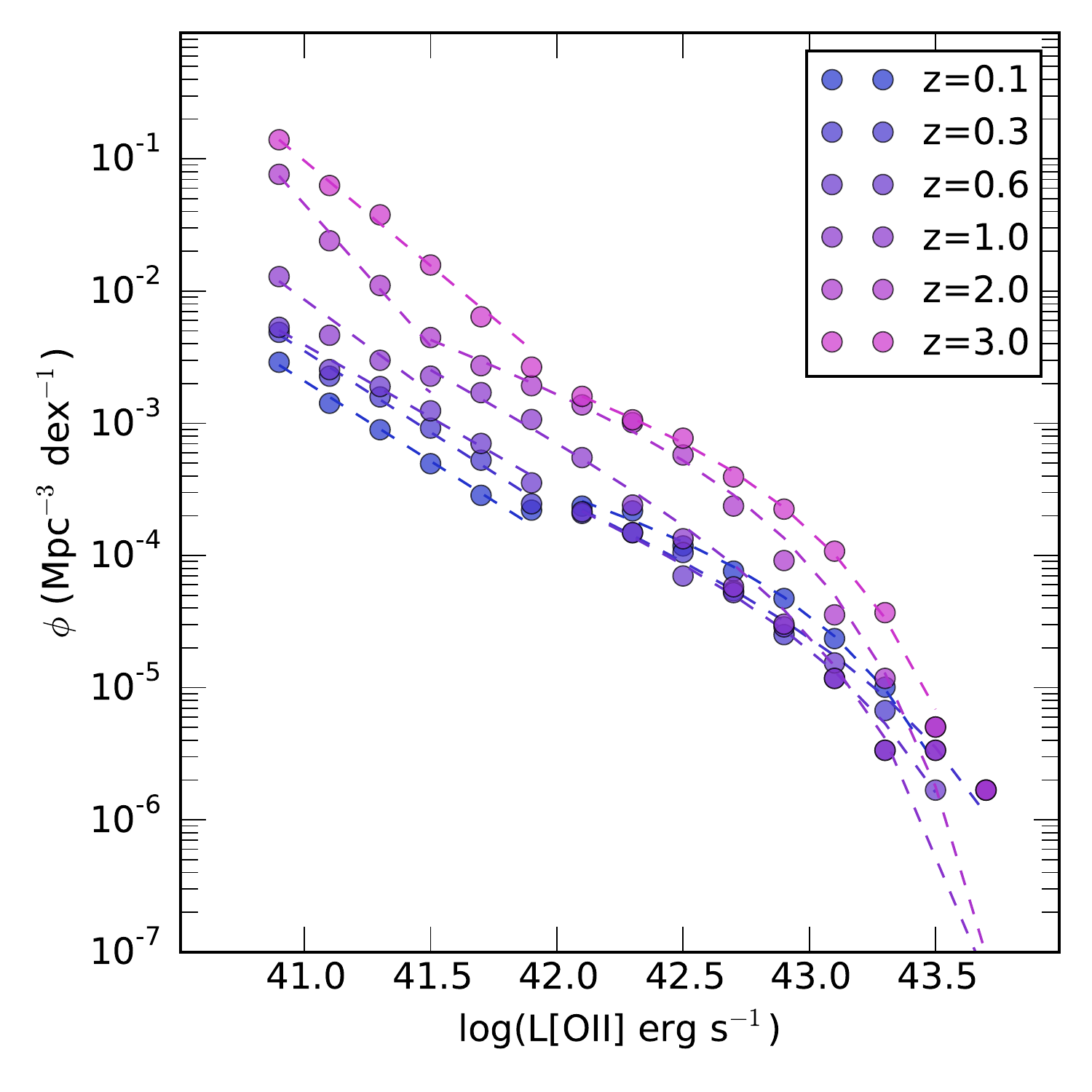}
\caption{Evolution of the LF of the \oii~emitting galaxies from
$z=3$ to $z=0.1$. In general, the LF increases and the slope of the
LF also becomes steeper with increasing redshift. There is strong
evolution of the LF for $z > 1.0$ but not below $z < 1.0$. }
\label{fig:evol_oii} 
\end{figure}

\begin{figure}
\includegraphics[width=85mm]{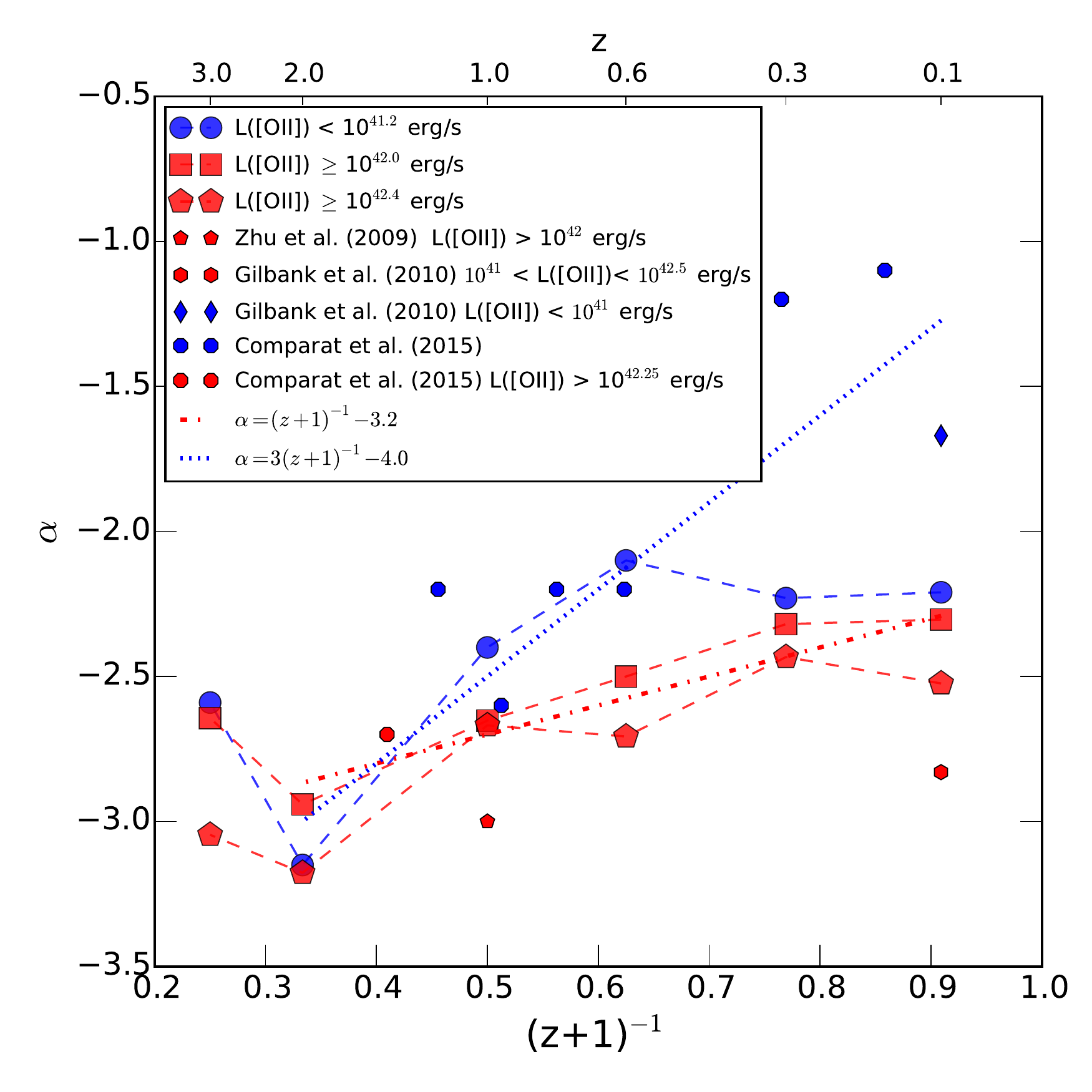}
\caption{The slope of the LFs of the \oii~emitting galaxies as a
function of redshift for the bright (squares) and faint (circles)
ends. For both ends, $\alpha$ shows a minor change at $z\le 0.3$.
For the bright end, $\alpha$ increases approximately as $(z+1)^{-1}$ (dot-dashed line)
while $\alpha$ of the faint end shows the bigger change being proportional to 
$\sim 3(z+1)^{-1}$ (dotted line) at high redshift $z\ge
0.6$. The slopes for both ends display the minima at $z=2.0$. }
\label{fig:oii_alpha} 
\end{figure}



\subsection{Fitting function of the [OII]~luminosity functions}
\begin{table*}
\centering
\caption{Fitting parameters of the LFs of [OII] emitters in equations~(\ref{eq:fit_lowl}) and (\ref{eq:fit_highl})}
\begin{tabular}{cccccccc}
\hline
z & $\alpha_l$ & $\beta_l$ & $\chi_{\rm l, red}^2$ & $\alpha_h$ & $\beta_h$ & $L_*$ & $\chi_{\rm h, red}^2$ \\
\hline
0.1 & -2.21 $\pm$  0.06  & 46.96 $\pm$  2.60 &  1.10 & -1.59 $\pm$  0.15  & -4.10 $\pm$  0.22 & 43.06 $\pm$  0.14 &  0.16 \\
0.3 & -2.23 $\pm$  0.07  & 47.88 $\pm$  2.90 &  2.16 & -1.91 $\pm$  0.13  & -4.80 $\pm$  0.34 & 43.40 $\pm$  0.21 &  0.14 \\
0.6 & -2.10 $\pm$  0.08  & 42.64 $\pm$  3.22 &  3.37 & -1.88 $\pm$  0.14  & -4.56 $\pm$  0.30 & 43.16 $\pm$  0.18 &  0.09 \\
1.0 & -2.40 $\pm$  0.26  & 55.37 $\pm$ 10.77 & 38.03 & -2.05 $\pm$  0.11  & -4.13 $\pm$  0.34 & 42.98 $\pm$  0.21 &  1.49 \\
2.0 & -3.15 $\pm$  0.14  & 86.61 $\pm$  5.92 & 37.80 & -1.74 $\pm$  0.06  & -3.35 $\pm$  0.14 & 42.85 $\pm$  0.09 &  1.14 \\
3.0 & -2.59 $\pm$  0.07  & 64.10 $\pm$  3.07 & 43.42 & -1.65 $\pm$  0.08  & -3.29 $\pm$  0.11 & 42.96 $\pm$  0.07 &  0.19 \\
\hline
\end{tabular}
\label{table:oii}
\end{table*}

Despite the fact that the \oii~LF is not in perfect
agreement with observations at low redshift, the \oii~LF at z=1
still shows a reasonably good agreement with observations. Therefore we fit
the \oii~LFs from the \mbii~for low redshift $z \le 1.0$, and extend
the fitting for LFs at high redshift in the range $1.0 < z \le 3.0$.
Since the exponential decline of the number density is obvious
toward the high luminosity end, we use two fitting
functions for lower \oii~luminosity than L(\oii) $= 10^{42}~$erg\,s$^{-1}$
(dotted vertical lines) 
and high luminosity (L(\oii) $> 10^{42}~$erg\,s$^{-1}$), respectively.
For $z =1.0$ and $2.0$, L(\oii) $= 10^{41.5}~$erg~s$^{-1}$ is used
instead as the border line to adopt the fact that the Schechter function
extends to lower luminosity. At low luminosity, single
power-law function is used while the common \citet{Schechter:1976}
function is used at high luminosity.

The Schechter function has the form
\begin{equation}
\phi (L)dL =\phi^* \left(\frac{L}{L^*} \right)^\alpha~ {\rm exp}
\left(-\frac{L}{L^*} \right) d \left(\frac{L}{L^*} \right) 
\end{equation}
where $\alpha$ is the slope of the faint end of the LF, $L^*$ is the characteristic
luminosity, and $\phi^*$ is the density of galaxies per magnitude with $L > L^*$. 
At low luminosity (L(\oii) $< 10^{42}$~erg\,s$^{-1}$) in each redshift,
a power-law fitting function is used with the form 
\begin{equation}
\phi(\logl)d \logl=10^{(\alpha_l +1)(\logl-42.0)+\beta_l} d \logl
\label{eq:fit_lowl}
\end{equation}
while the Schechter function is used for high luminosity as  
\begin{equation}
\phi(\logl)d\logl = 10^{(\alpha_h +1){\rm log}(\frac{L}{L^*})+{\beta_h}}~{\rm exp}(-\frac{L}{L^*}) d\logl.   
\label{eq:fit_highl}
\end{equation}

Fig.~\ref{fig:fit_oii} shows the LFs along with fitting parameters
at redshift $z=0.1, 0.3, 0.6, 1.0, 2.0$, and $3.0$, respectively.
We apply Poisson errors to estimate the uncertainty of the number
of galaxies in each luminosity bin (see Appendix~\ref{appendix:error}
for the comparison of the estimated Poisson and Jackknife errors).
Table~\ref{table:oii} shows the fitting parameters and the corresponding
reduced $\chi^2$ values at different redshifts. At high luminosity
ends, the number of galaxies are dominated by the Poisson errors.
For example, only 2--3 galaxies are found in the simulated volume
at the luminosity bin of L(\oii) $\sim 10^{43.6}$ erg\,s$^{-1}$.
However, at the low luminosity
ends, the small Poisson errors due to the high number of galaxies
(e.g., $\sim 73000$ galaxies at the luminosity bin L(\oii) $\sim
10^{40.8}$ erg\,s$^{-1}$ for $z=3.0$) return the high values of
$\chi^2$.

Fig.~\ref{fig:evol_oii} shows all the \oii~LFs from $z=3.0$ to the
local universe $z=0.1$ to show the LF evolution as a function of
redshift. In general, the number of galaxies increases and the slope
of the LFs decreases (i.e., $\alpha$ decreases) as a function of
redshift. The number of \oii~galaxies at the lowest luminosity increases
by almost 2 orders of magnitude from $z=0.1$ to $z=3.0$ whereas
relatively small change is observed at the highest luminosity
end. The LFs do not evolve much at low redshift $z<1.0$ but strong
evolution is observed at high redshift $1.0 \le z \le 3.0$.


\begin{figure*} \includegraphics[width=180mm]{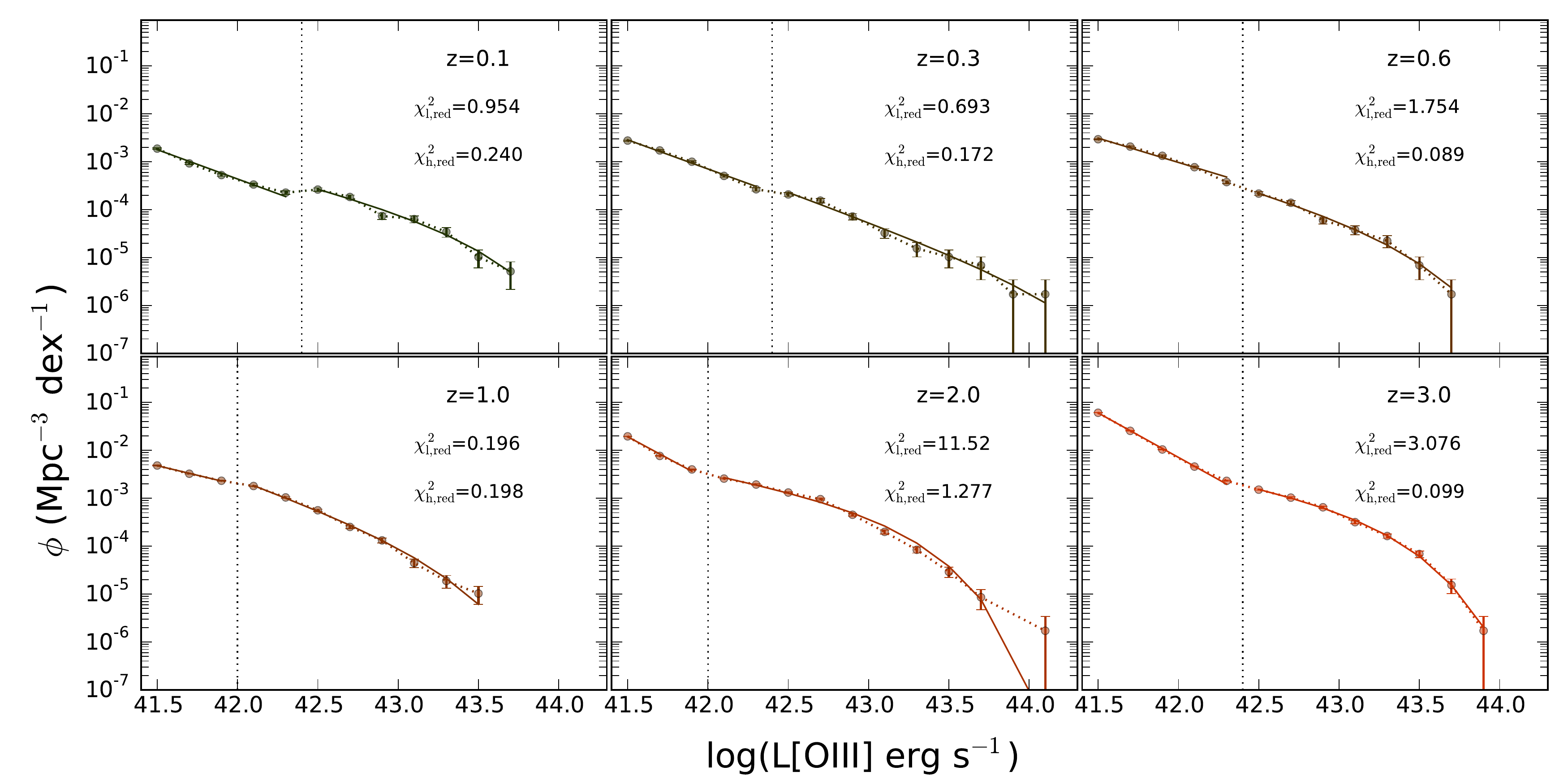}
\caption{LFs together with fitting functions of the \oiii~ELGs
at $z=0.1$, 0.3, 0.6, 1.0, 2.0, and 3.0. The \oiii~LF is
similar to the \oii~LF except for the fact that L(\oiii) is higher
than L(\oii) of a same SFR. The dotted vertical line at each redshift
shows the dividing luminosity for fitting LFs at low and
high luminosity.}
\label{fig:fit_oiii}
\end{figure*}

Fig.~\ref{fig:oii_alpha} shows the slope $\alpha$ of the LFs as a
function of redshift. The slopes for the bright (L(\oii) $\ge
10^{42.0}$ erg\,s$^{-1}$) and faint (L(\oii) $\le 10^{41.2}$
erg\,s$^{-1}$) ends of the LFs are plotted against $(z+1)^{-1}$ as
squares and circles, respectively. For both ends, $\alpha$ shows
a minor change at $z\le 0.3$.  For the bright end, the slope
changes approximately as $(z+1)^{-1}$ whereas the slope of the faint
end
shows the bigger change approximately being proportional to $\sim
3(z+1)^{-1}$ at high redshift $0.6 \le z\le 2.0$. The slopes for
both ends display the minima at $z=2.0$. Since the Poisson errors
dominate at the bright end of the LFs, the slopes for L(\oii) $\ge
10^{42.4}$ erg\,s$^{-1}$
are plotted additionally (pentagons) to see how the errors affect the
determination of the slopes. The number of \oii~galaxies at the
highest luminosity bins L(\oii) $\sim 10^{43.6}$ erg\,s$^{-1}$ is 
typically 2--3 in the simulated volume, and thus the Poisson
statistics greatly affects the determination of the slopes. For a
comparison with observations, the slopes from
\citet{Zhu:2009,Gilbank:2010,Comparat:2015} are plotted together
in Fig.~\ref{fig:oii_alpha}, however, note that their luminosity
cuts do not exactly match with the ones from the current work for
both bright and faint ends. For example, the luminosity
range of the bright end from \citet{Gilbank:2010} is 10$^{41.0} <$
L(\oii) $< 10^{42.5}~$erg\,s$^{-1}$. 
The evolution of the \oii~LF as a function of redshift
is driven by the evolution of GSMF, i.e., by the cosmic SFR in
different stellar masses. Thus, it is important to model the
SFR in various stellar mass ranges. At low mass end, the SFR is
greatly affected by the star-formation feedback model whereas the
AGN feedback greatly affects the high stellar mass end.

\subsection{[OIII]~luminosity functions}
\begin{figure}
\includegraphics[width=85mm]{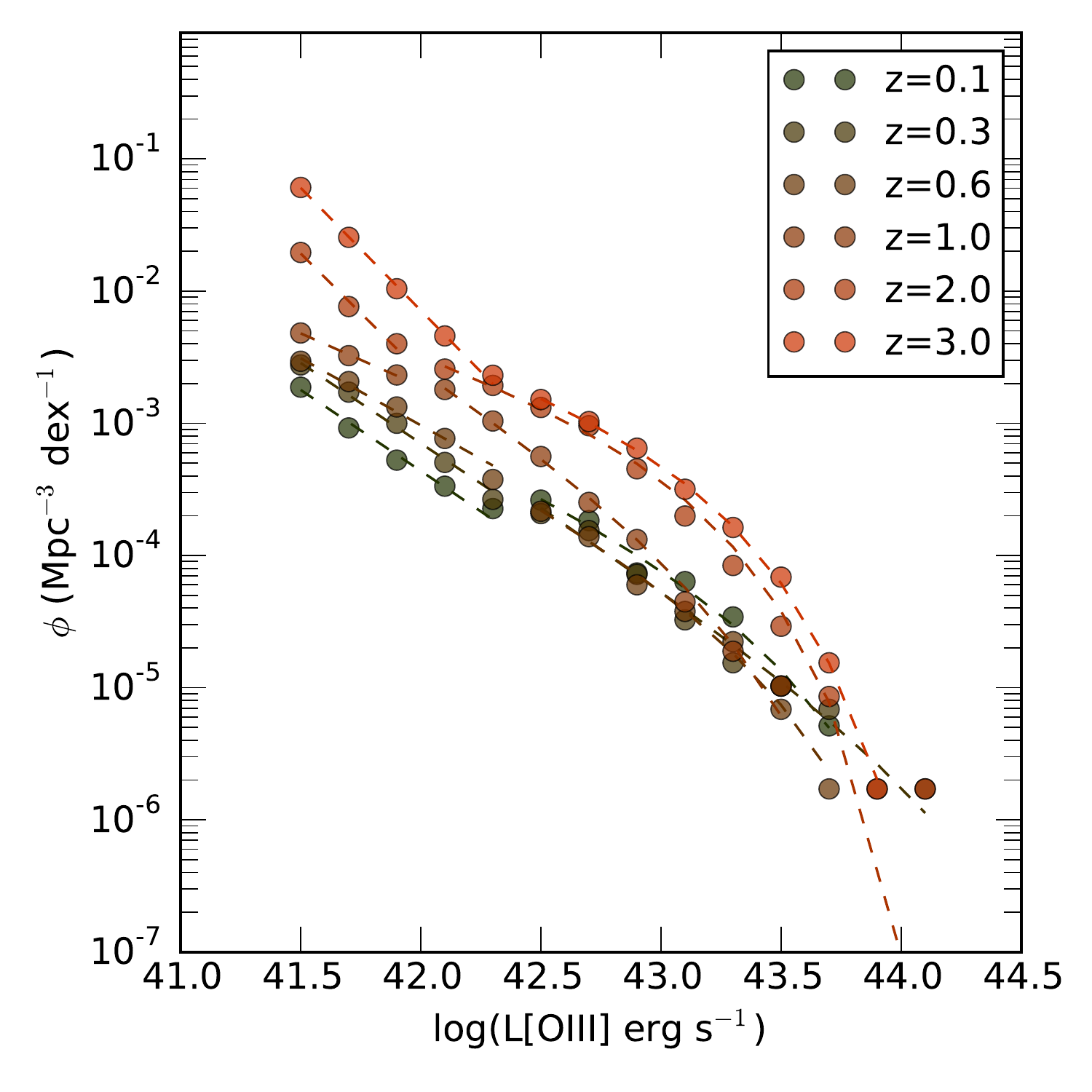}
\caption{Evolution of LF of the \oiii~emitting galaxies from $z=3$
to $z=0.1$. The overall shape of the \oiii~LF is similar to the one
for \oii~LF with a shift in luminosity.}
\label{fig:evol_oiii} 
\end{figure}

In this section, we briefly discuss the LFs of the \oiii~emitting
galaxies in the \mbii, since the upcoming survey \wfirst~will explore
the evolution of the \oiii~emission line LFs at high redshift. We
repeat the process for the LFs of the \oii~emitters for \oiii~emission
line. In the synthesised SEDs of the star-forming galaxies, the
emission-line fluxes of the \oii~and \oiii~lines are calculated in
a similar manner. Note that the \oiii~LFs in Fig.~\ref{fig:fit_oiii}
are very similar to the \oii~LFs in Fig.~\ref{fig:fit_oii} except
for the fact that the \oiii~luminosity is higher than \oii. We fit the
LFs with two functions as we do for \oii~LFs; however, we split the
LFs at the luminosity 0.4 dex higher than the luminosity for the
\oii~at L(\oiii) $= 10^{42.4}$~erg\,s$^{-1}$. Fig.~\ref{fig:fit_oiii}
shows the \oiii~LFs along with fitting functions at redshift $z=0.1,
0.3, 0.6, 1.0, 2.0$, and $3.0$, respectively. Fig.~\ref{fig:evol_oiii}
shows all the \oiii~LFs at $0.06 \le z \le3.0$ together; 
Table~\ref{table:oiii} lists the fitting parameters.

\begin{table*}
\centering
\caption{Fitting parameters of the LFs of [OIII] emitters in equations (\ref{eq:fit_lowl}) and (\ref{eq:fit_highl})}
\begin{tabular}{cccccccc}
\hline
z & $\alpha_l$ & $\beta_l$ & $\chi_{\rm l, red}^2$ & $\alpha_h$ & $\beta_h$ & $L^*$ & $\chi_{\rm h, red}^2$ \\
\hline
0.1 & -2.23 $\pm$  0.08  & 48.10 $\pm$  3.51 &  0.95 & -1.90 $\pm$  0.26  & -4.43 $\pm$  0.55 & 43.50 $\pm$  0.33 &  0.24 \\
0.3 & -2.22 $\pm$  0.06  & 47.88 $\pm$  2.51 &  0.69 & -2.24 $\pm$  0.16  & -5.76 $\pm$  1.13 & 44.22 $\pm$  0.71 &  0.17 \\
0.6 & -2.01 $\pm$  0.08  & 39.57 $\pm$  3.52 &  1.75 & -2.01 $\pm$  0.20  & -4.55 $\pm$  0.44 & 43.43 $\pm$  0.24 &  0.09 \\
1.0 & -1.80 $\pm$  0.04  & 30.96 $\pm$  1.49 &  0.20 & -2.22 $\pm$  0.08  & -4.09 $\pm$  0.25 & 43.23 $\pm$  0.13 &  0.20 \\
2.0 & -2.80 $\pm$  0.18  & 73.08 $\pm$  7.36 & 11.52 & -1.67 $\pm$  0.10  & -3.24 $\pm$  0.16 & 43.16 $\pm$  0.10 &  1.28 \\
3.0 & -2.86 $\pm$  0.04  & 75.79 $\pm$  1.52 &  3.08 & -1.67 $\pm$  0.07  & -3.23 $\pm$  0.08 & 43.23 $\pm$  0.05 &  0.10 \\
\hline
\end{tabular}
\label{table:oiii}
\end{table*}


\subsection{Evolution of the auto-correlation function of [OII]~emitters}
\begin{figure}
\includegraphics[width=85mm]{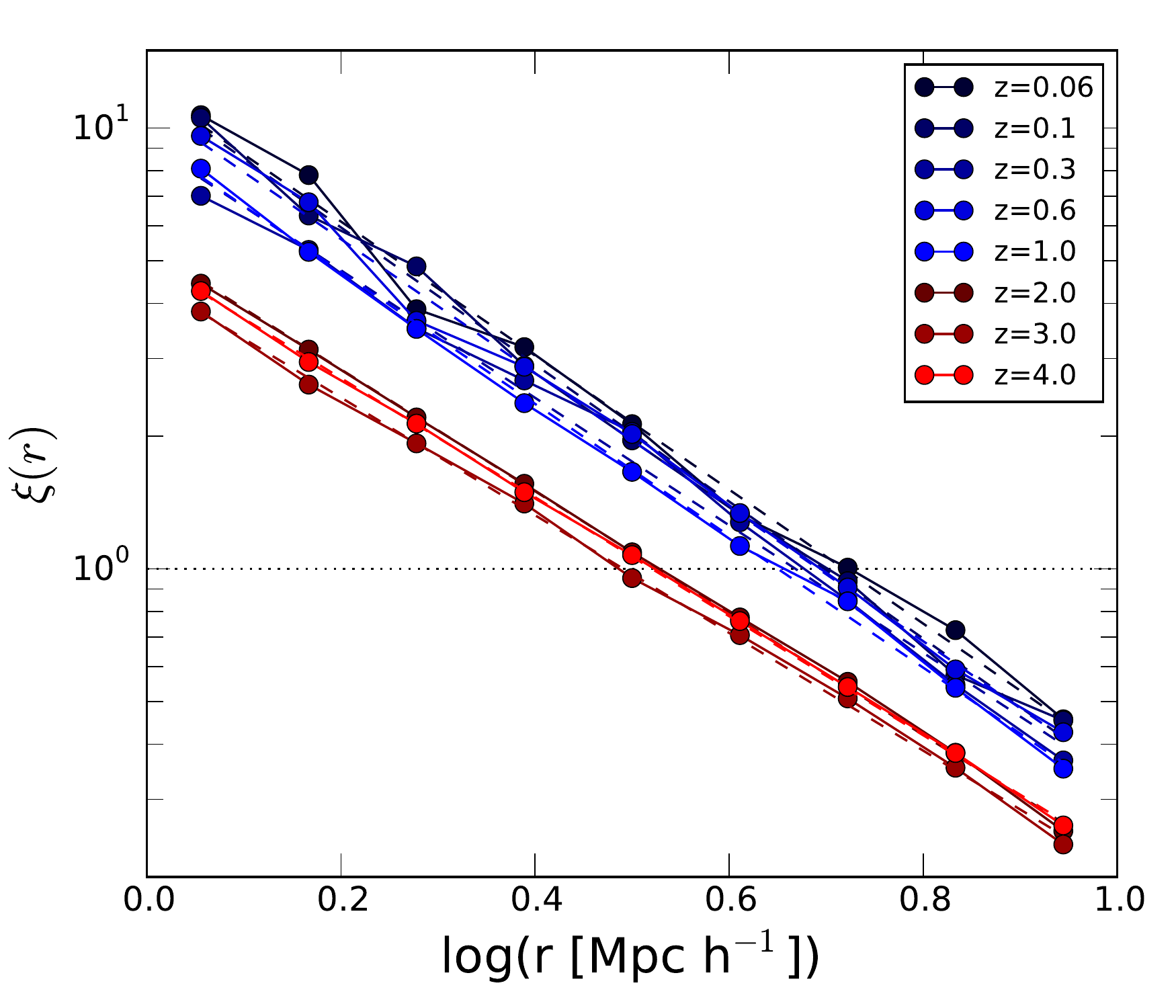}
\caption{The auto-correlation function [$\xi$(r) =DD(r)/RR(r)
-1] of the \oii~emitting galaxies for redshift $z=$~0.06, 0.1,
0.3, 0.6, 1.0, 2.0, 3.0, and 4.0. $\xi$ shows a significant evolution
from $z=2.0$ to $1.0$. The intersection between the
power-law fitting function $\xi_{\rm fit} (r)$
(dashed) and $\xi=1$ (dotted) shows the correlation length $l_{\rm
corr}$ for each redshift.}
\label{fig:auto_corr} 
\end{figure}

In this section, using the spatial distribution of the \oii~emission-line
selected galaxies, we explore the evolution of the 2-point
auto-correlation function [$\xi$(r)
=DD(r)/RR(r) -1] at $0.06 \le z \le 4.0$, 
where DD(r) is the number of galaxy pairs with separation r
and RR(r) is the number of pairs with the same separation for a
random (i.e., Poisson) distribution.
Fig.~\ref{fig:auto_corr}
shows the auto-correlation functions of \oii~emitters at the redshifts
$z=0.06, 0.1, 0.3, 0.6, 1.0, 2.0, 3.0,$ and $4.0$ along with the corresponding
fitting functions shown as dashed lines. 
When the auto-correlation function is fitted with a power-law as $\xi_{\rm fit}
(r) = (r/l_{\rm corr})^{m_{\rm corr}}$, $\xi_{\rm fit} (l_{\rm corr})  = 1$ is
obtained where $l_{\rm corr}$ is the correlation length within which
the galaxy distribution is correlated. In Fig.~\ref{fig:auto_corr},
the intersection between the power-law fitting function for each
redshift $\xi_{\rm fit} (r)$ and $\xi=1$ shows the correlation
length $l_{\rm corr}$ for each redshift. 
Table~\ref{table:fitting_auto}
lists the fitting parameters of the auto-correlation functions.
The $\xi$ shows a significant evolution from $z=2.0$ to 1.0 whereas
it evolves mildly at $z \le 1.0$ or $z \ge 2.0$. At $z \le 1.0$, the
slopes remain constant in the range $m_{\rm corr}=-1.55$ to $-1.45$
while the correlation length 
increases from $l_{\rm corr} = 4.47$~Mpc
$h^{-1}$ at $z=1.0$ to $l_{\rm corr} = 5.23$~Mpc\,$h^{-1}$ at
$z=0.06$. At $z \ge 2.0$, the slopes display slightly larger values
than the ones for $z \le 1.0$ in the range of $m_{\rm corr}=-1.38$
to $-1.33$ and the smaller correlation lengths are clearly observed
in the range $l_{\rm corr} = 3.10$--$3.37$~Mpc\,$h^{-1}$.



\begin{table}
\centering
\caption{Fitting parameters of the auto-correlation functions}


\begin{tabular}{ccc}
\hline
z &  $m_{\rm corr}$ & $l_{\rm corr}$\,(Mpc $~h^{-1}$) \\
\hline
0.06   & -1.52 &    5.23  \\
0.1    & -1.55 &    4.99  \\
0.3    & -1.45 &    4.66  \\
0.6    & -1.52 &    4.92  \\
1.0    & -1.50 &    4.47  \\
2.0    & -1.38 &    3.37  \\
3.0    & -1.33 &    3.10  \\
4.0    & -1.35 &    3.31  \\
\hline
\end{tabular}
\label{table:fitting_auto}
\end{table}

\section{Summary and discussion}
\label{sec:summary}
In this paper, we investigate the properties of the \oii~ELGs in
the state-of-the-art high-resolution cosmological simulation \mbii.
The \mbii~simulates galaxies including baryonic physics for star-formation
and AGN feedback in a comoving volume of $100~h^{-1}$~Mpc
on a side from $z=159$ to $z = 0.06$. From the synthesised SEDs
of the individual galaxies, which includes the stellar continuum and
the emission lines based on the star formation history, we select
a sub-sample of star-forming galaxies based on the 
\oii~luminosity in the redshift range  $0.06 \le z \le 3.0$. We validate the use of the
extracted L(\oii) by comparing it with several observations. The
\mbii~simulation shows a good agreement with observations, and we
focus on the evolution of the \oii~(and \oiii) LFs in the redshift
range $0.06 \le z \le 3.0$. We expect that the current work is
useful for forthcoming surveys such as \hetdex, \desi, and \wfirst.
We summarise our main findings as follows.

\begin{itemize}

\item The specific SFR as a function of stellar
mass agrees with the \gama~survey when galaxies with high
\oii~luminosity are selected as L(\oii) $\ge 10^{41.5}~$erg\,s$^{-1}$.
\\

\item We show that the \oii~LF at $z = 1.0$ from the \mbii~shows a
good agreement with the LFs from several surveys \citep{Zhu:2009,
Comparat:2015} below L(\oii)=$10^{43.0}$ erg\,s$^{-1}$ while the
low redshifts ($z \le 0.3$) show an excess in the prediction of
bright \oii~galaxies, but still displaying a good match with
observations \citep{Gilbank:2010,Ciardullo:2013, Comparat:2015}
below L(\oii)=$10^{41.6}$ erg\,s$^{-1}$. \\

\item We present the LFs of \oii~and \oiii~ELGs at the redshift
range $0.1 \le z \le 3.0$ and provide fitting functions for each
redshift. Each \oii~LF at different redshift is fitted with a
single power-law at low luminosity (L(\oii) $\la 10^{42.0}~$erg\,s$^{-1}$)
while Schechter function is applied at high luminosity (L(\oii) $\ga 
10^{42.0}~$erg\,s$^{-1}$).  In general, the LF increases and the
slope of the LF also becomes steeper with increasing redshift.  The
slopes of the LFs at bright and faint ends range from -3 to -2
showing minima at $z=2$. The slope of the bright end evolves
approximately as $(z+1)^{-1}$ at $z \le 2$ while the faint end
evolves as $\sim 3(z+1)^{-1}$ at $0.6 \le z \le 2$.  We apply a
similar analysis to the \oiii~LF with a shift of 0.4 dex. \\ 


\item The auto-correlation function of [OII] ELGs shows
a significant evolution from $z=2$ to $1$ while it changes 
mildly at $z \le 1$ or  $2.0 \le z \le 4.0$. The correlation length
increases from $\sim$3 Mpc\,$h^{-1}$ for $z\ge 2$ to
$\sim$5 Mpc\,$h^{-1}$ for $z\le 1$.  

\end{itemize}

The current theoretical study of the LFs of \oii~(and \oiii) emitting
galaxies will be useful for the forthcoming surveys \hetdex,
\desi, and \wfirst. The current high resolution of \mbii~simulations
makes it possible to investigate the \oii~line of an individual galaxy
and construct a better theoretical model for LFs, but the low
luminosity end still suffers from shot noise. We expect that 
future simulations with even higher resolution can provide a solid
feature at low luminosity. At the bright end of the LFs where 
Poisson statistics dominates, enlarging the simulation volume will 
help to get better errors for each luminosity bin.

The excess of galaxies at high luminosity end in
\oii~(and \oiii) LF is inevitably transferred from the overabundance
at the high mass end in GSMF of the \mbii. In this respect, the future
theoretical study of \oii~LF should be focused on the GSMF by
applying observation matching star-formation and AGN feedback models. Especially
at the faint end, a stronger
star-formation feedback model \citep[e.g.,][]{Okamoto:2010} used
in \textsc{Illustris} \citep{Vogelsberger:2014} and \textsc{BlueTides}
\citep{Feng:2015} can reduce the number of star-forming galaxies.
On the bright end, increasing the AGN feedback \citep[e.g.,
\textsc{eagle} simulation by][]{Schaye:2015} can suppress the
star-formation by driving the gas out
of the halos.  

The dust obscuration in star-forming galaxies in low and high
redshift galaxies remains as an open question. Future studies should
consider the effect of dust reddening on the LF at different
redshifts.

\section*{Acknowledgements} KP is supported by the Urania E. Stott
Fellowship of The Pittsburgh Foundation and partly by the National
Science Foundation under the Theoretical and Computational Astrophysics
Network (TCAN) grant AST-1333360. TDM acknowledges the National
Science Foundation, NSF Petapps, OCI-0749212 and NSF AST-1009781
for support. SH is supported by NSF-AST1412966, NASA-NNH12ZDA001N-
EUCLID and DOE-DESC0011114. Authors thank the anonymous
referee for constructive comments. Authors also thank John Wise, Edmund
Hodges-Kluck, and JongHak Woo for useful comments. The numerical
analysis presented in this paper was performed using the cluster
facilities (``ferrari" and ``coma") of the McWilliams Center for
Cosmology at Carnegie Mellon University.

\bibliographystyle{mn2e} 
\bibliography{park_sf}


\appendix
\section[]{Validation of the [OII] LF}
\label{appendix:lf_raw}
\begin{figure}
\includegraphics[width=85mm]{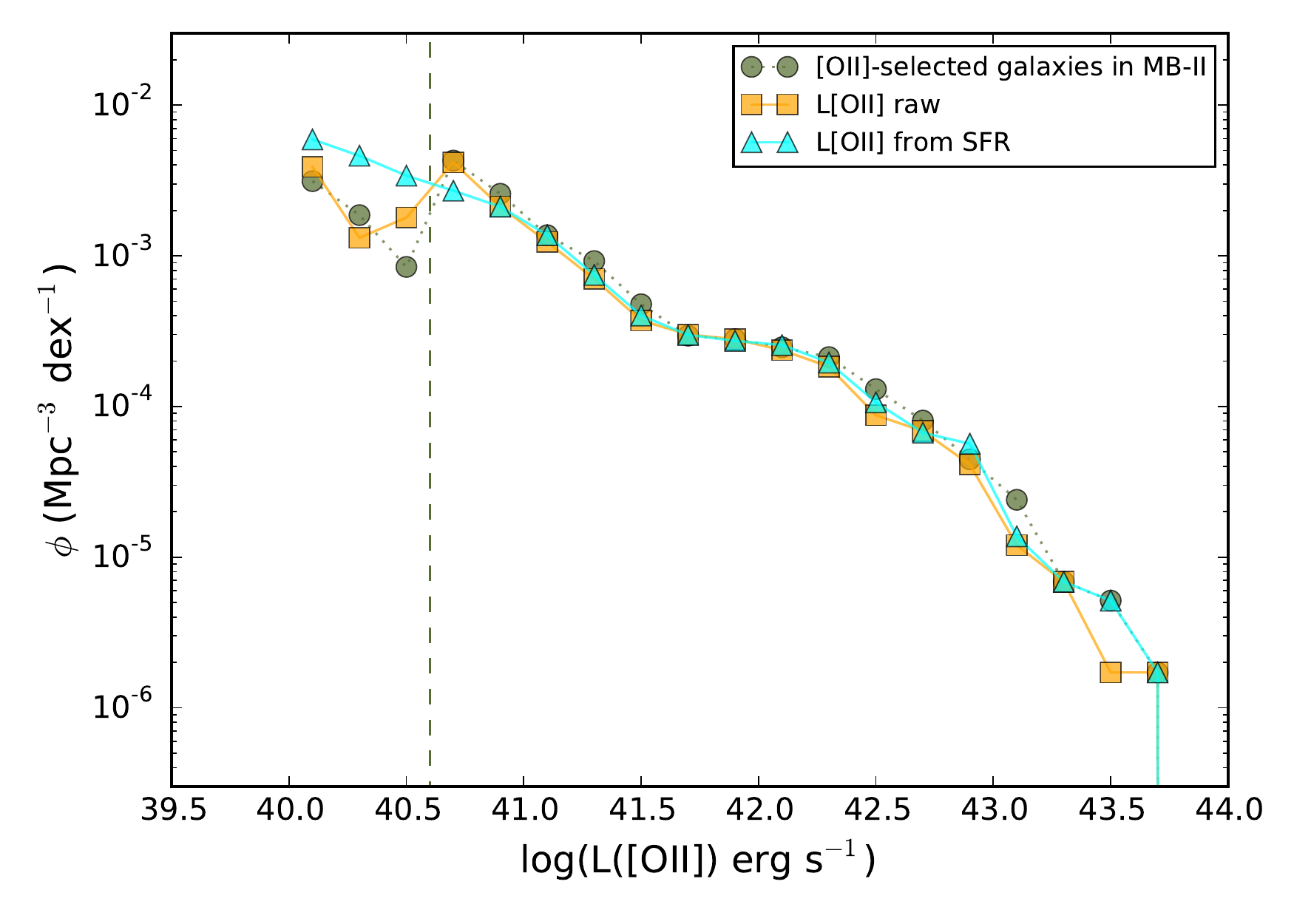}
\caption{LFs of \oii~emitting galaxies in the \mbii~at $z=0.06$.
Circles show the \oii-selected galaxies while squares show the LF
of the raw L(\oii) before the emission lines are added to the stellar
SEDs for each galaxy. Triangles show the LF of the L(\oii) which
is converted from the SFR using the empirical relationship by
\citet{KewleyGJ:2004}. Our method of selecting \oii~ELGs
recovers the LF of the intrinsic \oii~lines well. The LF
starts to lose star-forming galaxies due to the fact that the
\oii~lines trace the average SF history of the past $\sim 20$~Myr.
Note that the LF converted from the SFRs keeps increasing toward
the low luminosity end. Shot noise due to the mass of the star
particles also starts to dominate in the low luminosity L(\oii) $<
10^{40.6}$erg\,s$^{-1}$ shown as a vertical dashed line.} 
\label{fig:lf_raw} \end{figure}

We compare the \oii~LFs calculated using different methods for the
purpose of validation in Fig.~\ref{fig:lf_raw}. We confirm that our
method recovers the intrinsic LF of the \oii~well, as shown in
Fig~\ref{fig:lf_raw}. The sampling of the \oii~emitting galaxies
start to lose star-forming galaxies at L(\oii) $\le 10^{40.6}$erg\,s$^{-1}$
due to the sensitivity of the \oii~emission line on the SF timescale
and the shot noise due to the mass resolution of the gas particles
in the \mbii~simulation.

\begin{figure}
\includegraphics[width=85mm]{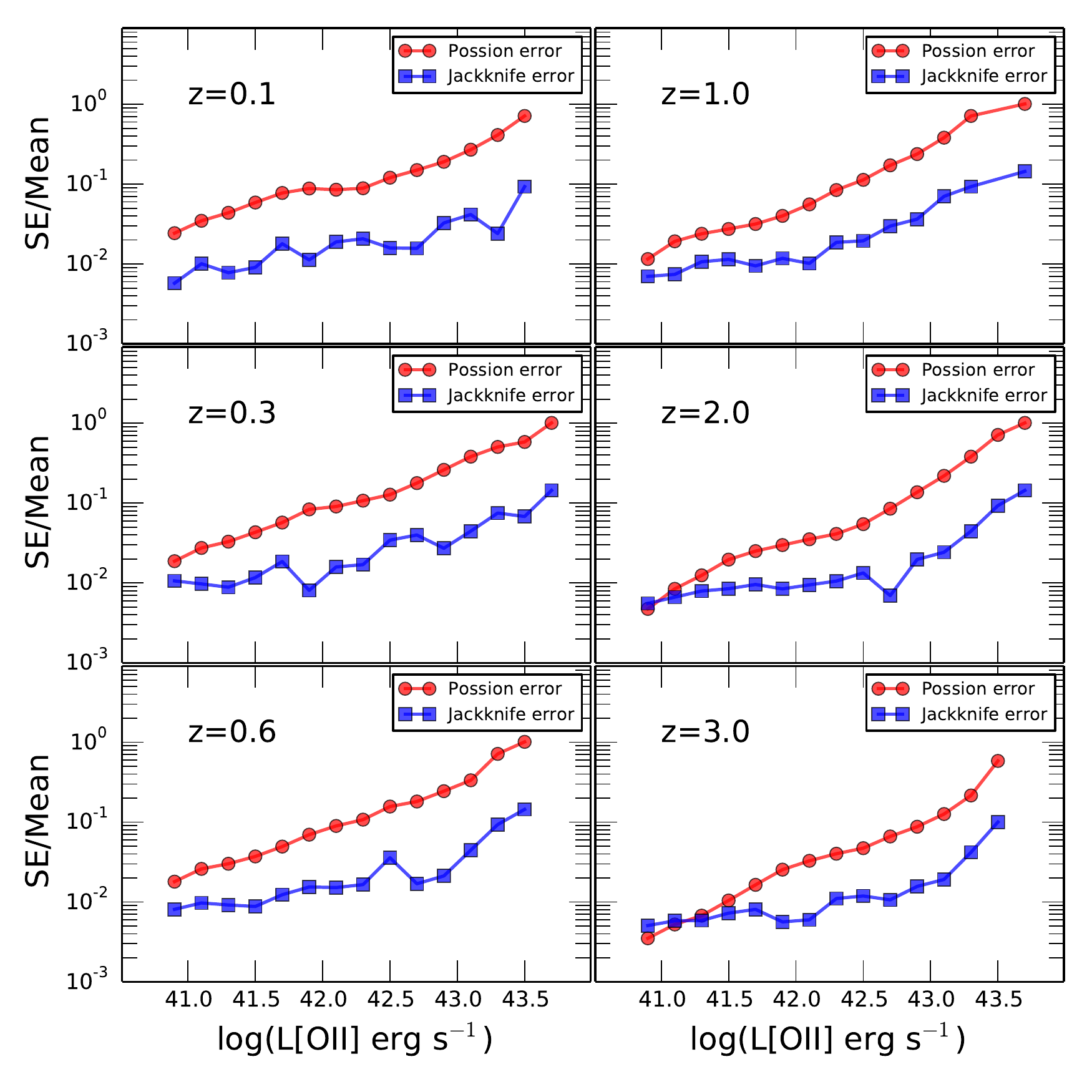}
\caption{Comparison of Poisson and Jackknife re-sampling errors.
For each luminosity bin, Poisson errors ($\sqrt{N}$) are obtained
(circles) from the number of galaxies ($N$). Jackknife sampling of
8 cubes with comoving volume of $50$~Mpc~$h^{-1}$ on a side is
used to get errors due to the cosmic variance. The Poisson
and Jackknife errors, shown here, are normalised by the mean value.}
\label{fig:errors}
\end{figure}

\section[]{Standard errors of the LF}
\label{appendix:error}
We compare the errors of the LFs in each redshift from Poisson
statistics and Jackknife re-sampling. Fig.~\ref{fig:errors} shows
the standard errors normalised by the mean values for each luminosity
bin. Poisson errors ($\sqrt{N}$) are obtained (circles) from the
number of galaxies ($N$) in each luminosity bin. Jackknife sampling
of 8 cubes with comoving volume of ($50$~Mpc~$h^{-1}$)$^3$ is used
to get the errors to consider the cosmic variance. Poisson errors
are bigger than the Jackknife re-sampling errors, whereas both errors
are comparable at low luminosity L(\oii) $\sim 10^{41}$~erg\,s$^{-1}$.

\label{lastpage}
\end{document}